\documentclass[conference]{IEEEtran}
\IEEEoverridecommandlockouts
\usepackage[noadjust]{cite}
\usepackage{amsmath,amssymb,amsfonts}
\usepackage[bookmarks=false]{hyperref}
\hypersetup{draft}

\usepackage{graphicx}
\usepackage{textcomp}
\usepackage{xcolor}
\newcommand{\etal}{\textit{et al. }}

\def\BibTeX{{\rm B\kern-.05em{\sc i\kern-.025em b}\kern-.08em T\kern-.1667em\lower.7ex\hbox{E}\kern-.125emX}}

\pdfoutput=1
\usepackage{caption}
\usepackage{amsmath}
\usepackage{amssymb}
\usepackage{graphicx}
\usepackage{enumitem}

\usepackage{algorithmicx}
\usepackage[ruled,linesnumbered]{algorithm2e}
\usepackage{algpseudocode}

\usepackage{flushend}

\newcommand{\removelatexerror}{\let\@latex@error\@gobble}
\usepackage{multirow}
\usepackage{graphicx}
\usepackage{subfigure}
\usepackage[utf8]{inputenc}
\usepackage[english]{babel}
\usepackage{amsthm}

\IEEEsettopmargin{t}{0.80in}
\usepackage{booktabs}
\begin{document}
\title{SearchFromFree: Adversarial Measurements for Machine Learning-based Energy Theft Detection}

\author{
    \IEEEauthorblockN{Jiangnan Li$^{\ast}$, Yingyuan Yang$^{\S}$, Jinyuan Stella Sun$^{\ast}$}
    \IEEEauthorblockA{$^{\ast}$The University of Tennessee, Knoxville, {\{jli103, jysun\}@utk.edu}
    \\$^{\S}$University of Illinois Springfield, yyang260@uis.edu}
}

\IEEEoverridecommandlockouts
\IEEEpubid{\makebox[\columnwidth]{\textbf{This paper has been accepted by IEEE SmartGridComm 2020.} \hfill} \hspace{\columnsep}\makebox[\columnwidth]{ }}
\maketitle
\IEEEpubidadjcol
\begin{abstract}

Energy theft causes large economic losses to utility companies around the world. In recent years, energy theft detection approaches based on machine learning (ML) techniques, especially neural networks, become popular in the research literature and achieve state-of-the-art detection performance. However, in this work, we demonstrate that the well-perform ML models for energy theft detection are highly vulnerable to adversarial attacks. In particular, we design an adversarial measurement generation algorithm that enables the attacker to report extremely low power consumption measurements to the utilities while bypassing the ML energy theft detection. We evaluate our approach with three kinds of neural networks based on a real-world smart meter dataset. The evaluation result demonstrates that our approach can significantly decrease the ML models' detection accuracy, even for black-box attackers.

\end{abstract}


\section{Introduction}

As one of the primary non-technical losses (NTL) in the power systems, energy theft happens when an attacker deliberately manipulates his/her electricity data to reduce the electricity bills. To date, energy theft causes high financial losses to electric utility companies around the world. It is estimated that up to \$6 billion of electricity is stolen (around 1\% to 3\% annual revenue) in the United States each year \cite{forbesNews}, and the loss can be larger in developing countries \cite{gao2019physically}. 

In recent years, with the development of advanced metering infrastructure (AMI), two-way data communications between customers and utilities are enabled by the smart meters. However, recent studies show that the AMI introduces new vulnerabilities that can be exploited by the attackers to launch energy theft \cite{zanetti2017tunable}. Meanwhile, the smart meters are shown to be vulnerable to physical penetration \cite{hackmeter}, and there are even video tutorials online on smart meter hacking \cite{youtubeHacking}. Therefore, the energy theft problem is serious and the corresponding detection approaches are needed.

Different approaches were proposed to detect and mitigate energy theft. In general, the detection approaches can be categorized into two classes, sensor-based and consumption profile-based. The sensor-based approaches require the deployment of extra equipment and will increase the maintenance cost while the profile-based approaches utilize the customers' power usage patterns to detect abnormal variations \cite{zanetti2017tunable}. Among the profile-based approaches, detection methods that employ machine learning (ML) techniques, especially deep neural networks (DNN), are becoming popular in the research community \cite{zanetti2017tunable}\cite{nabil2018deep, jokar2015electricity, korba2018energy, zheng2017wide}. Enabled by the AMI, ML-based approaches take advantage of the statistical properties of the massive fine-grained smart meter data and achieve state-of-the-art detection performances. Meanwhile, the ML detection schemes are purely data-driven without additional equipment, which makes them compatible with current infrastructures in the power systems.

However, recent studies in the computer vision domain have demonstrated that ML models are highly vulnerable to adversarial attacks \cite{szegedy2013intriguing, goodfellow2014explaining, rozsa2016adversarial, moosavi2016deepfool}.  Meanwhile, the adversarial attacks are also shown to be effective in power system applications \cite{li2020conaml}\cite{chen2018machine}. As the ML approaches become popular in detecting energy theft, the threat from adversarial attacks needs to be investigated to prevent potential financial losses. 

In general, the process to generate adversarial examples can be represented as an optimization problem. The adversarial perturbations in the computer vision domain are expected to be small enough to avoid being noticed by human eyes. However, such a constraint is not compatible with a `smart' energy thief.  In particular, the attacker needs to generate a small energy consumption measurement to reduce his/her bill while bypassing the utilities' energy theft detection.  

In this work, from the attacker's point of view, we study the reliability of ML models used for energy theft detection. We analyze and formulate the properties of adversarial attacks in energy theft detection and propose a general threat model. Meanwhile, we design and implement an adversarial measurement generation algorithm that maximizes the attacker's profit. The evaluation of a real-world smart meter dataset shows that our algorithm can generate extremely low fake measurements while bypassing the well-trained ML detection models, even for black-box attackers. The main contribution of this paper can be summarized as follows:

\begin{itemize}

\item We summarize the properties of adversarial measurements in energy theft detection and propose a general threat model.

\item We design \textit{SearchFromFree}, an adversarial measurement generation algorithm to maximize the attackers' profit.

\item We evaluate our algorithm with three kinds of neural networks that are trained with a real-world smart meter dataset. The evaluation results demonstrate that our algorithm generates extremely low fake measurements that can still successfully bypass the ML models' detection. We open-source the related source code on Github \cite{SearchFromFree}.

\end{itemize}

Related research is discussed in section \ref{sec:related}. We format the attack and present the threat model in section \ref{sec:formation}. Section \ref{sec:design} and \ref{sec:implementation} present the algorithm design and the experiment evaluations respectively. We discuss the future work in section \ref{sec:discussion}. Finally, section \ref{sec:conclusion} concludes the paper.

\section{Related Work} \label{sec:related}

\subsection{ML-based Energy Theft Detection}

In 2009, Nagi \etal utilized the support vector machine (SVM) to detect abnormal power usage behaviors based on the historian consumption data \cite{nagi2009nontechnical}. After that, Depuru \etal extended their approach and introduced more information, such as the type of consumer, geographic location, to train the SVM classifier \cite{depuru2011support}. Meanwhile, SVM can also be combined with other techniques, such as fuzzy inference system \cite{nagi2011improving} and decision tree \cite{jindal2016decision}, to increase the detection performance. In 2015, Jokar \etal generated a synthetic attack dataset and trained a multiclass SVM classifier for each customer to detect malicious power consumption \cite{jokar2015electricity}. Recently, detection approaches based on DNNs have become popular and achieved promising performance. In 2017, Zheng \etal \cite{zheng2017wide} firstly employed deep convolutional neural networks (CNN) trained by a dataset released by State Grid Corporation of China (SGCC) and achieved a high detection accuracy with a low false-negative rate. After that, in 2018, a deep recurrent neural network (RNN) was employed in \cite{nabil2018deep} for energy theft detection. They trained the RNN with the daily smart meter data and randomly searched for the appropriate model parameters. In addition to the consumption domain, attackers may also claim higher supplied energy and overcharge the utilities in the smart grid to make a profit, such as through the photovoltaic solar cells. \cite{ismail2020deep} investigated the energy theft attack in the distributed generation domain and utilized a hybrid neural network to detect such attacks. 

\subsection{Adversarial Attacks}

The effect of the adversarial examples of neural networks was firstly discovered in the computer vision domain by Szegedy \etal in 2013 \cite{szegedy2013intriguing}. After that, different perturbation generation algorithms were proposed, such as the Fast Gradient Sign Method by Goodfellow \etal \cite{goodfellow2014explaining}, Fast Gradient Method by Rozsa \etal \cite{rozsa2016adversarial}, and DeepFool by Moosavi-Dezfooli \etal \cite{moosavi2016deepfool}. In recent years, the threat of adversarial attacks to power system applications also drew more and more attention. In 2018, Chen \etal demonstrated that adversarial examples are effective in both classification and regression ML applications in the power system \cite{chen2018machine}. In 2020, \cite{li2020conaml} demonstrated that the attacker can easily bypass the ML-based false data injection detection in power grid state estimation under the physical constraints. In \cite{marulli2019adversarial}, Marulli \etal studied the data poisoning attacks to ML models in energy theft detection using the generative adversarial network (GAN). However, they did not consider the evasion attacks and the profit of the attackers.

\section{Attack Formation} \label{sec:formation}

\subsection{Mathematical Presentation}

In this paper, we consider the scenario that the utilities employ a ML model $f_{\theta}$ to detect energy theft based on the customers' power consumption measurements. The model $f_{\theta}: M \rightarrow Y$ maps the measurement vector $M$ to its label $Y$ (Normal, Malicious) and is trained with the dataset $\left \{ M, Y \right \}$. The energy thief is assumed to be able to compromise his/her electricity meter and freely modify the meter's measurements. The purpose of the energy thief is to launch a false-negative evasion adversarial attack to $f_{\theta}$ by generating malicious measurement vectors $A$ so that $f_{\theta}(A) \rightarrow Normal$. Different from the general adversarial attacks which require the adversarial perturbations to be small, in the energy theft problem, the attacker only needs to focus on the size of $A$ in order to minimize his/her electricity bills. Therefore, the adversarial attacks in energy theft detection can be formally represented as follow:

\begin{subequations}
	\begin{align}
	\min \;\;\;& \left \| a \right \|_{1}\\
	s.t. \;\;\; & f_{\theta}(a) \rightarrow Normal \\ 
	& a_{i} \geq 0
	\end{align}
\end{subequations}

where $a$ represents a specific adversarial measurement vector, $\left \| a \right \|_{1} = \sum a_{i} $ is the $L_1$-Norm of $a$. The constraint (1c) requires that all the power consumption measurement $a_{i}$ in $a$ must be non-negative to be feasible.

\subsection{Threat Model}

We propose a practical threat model for the adversarial attacks in energy theft detection.

\begin{itemize}

\item We assume that the attacker is able to compromise his/her electricity meter and can freely modify the meter measurements. In practice, this can be implemented by physical penetration.

\item We consider a black-box adversarial attack since the energy thief usually cannot access the ML model $f_{\theta}$ and training dataset $\left \{ M, Y \right \}$ that are owned by the utilities. However, we allow the attacker to obtain an alternative dataset $\left \{ M', Y' \right \}$, such as a historian dataset, to train his/her model $f'_{\theta'}$ to generate adversarial measurements.

\item As demonstrated by (1c), the attacker needs to generate non-negative adversarial measurements in order to be practical.

\end{itemize}

In addition to the black-box attack described above, we will also evaluate the attack performance in the white-box scenario that allows the attacker to fully access the ML model $f_{\theta}$. Similar to the Kerckhoffs's principle in cryptography, such evaluations can demonstrate the vulnerability of the detection system under the worst-case scenario and allow us to learn the upper bound performance of the adversarial attacks.

\section{SearchFromFree Algorithm Design} \label{sec:design}

Based on our threat model, we propose the design of the \textit{SearchFromFree} algorithm to generate valid adversarial measurement vectors $A$ that can deceive the ML model $f_{\theta}$. Our algorithm assumes the attacker trained his/her local ML model $f'_{\theta'}$ with an alternative dataset, such as a public historian dataset. The assumption behinds our approach is the transferability of adversarial examples. Similar to the \textit{DeepFool} algorithm \cite{moosavi2016deepfool}, our algorithm is gradient-based and will iteratively search for a valid $a$.

\begin{algorithm}
\SetAlgoLined
\textbf{Input:} $f' _{\theta'}$, $step$, $size$, $\lambda$, $\sigma$ \\
\textbf{Output:} $a$\\

\SetKwBlock{Begin}{function}{end function}
\Begin($\text{advGen} {(} f' _{\theta'}, step, size, \lambda, \sigma {)}$)
{
  \textbf{define} $L_{total} = L(f'_{\theta'}(a), Y_{a}) - \lambda \cdot \left \| a \right \|_{1}$ \\
  initialize $a \sim N(0, \sigma^{2} )$ \\
  set all negative elements in $a$ to zero \\
  initialize $stepNum = 0$ \\
  
  \While{$stepNum \leq step - 1$ }{
  \If{$f'_{\theta'}(a) \rightarrow Normal$}{
    \Return{$a$}
  } 
  
  calculate gradient $ G = \nabla_{a} L_{total}$\\

  $r = G \cdot size / \textbf{max}(\textbf{abs}(G))$ \\
  update $a = a + r$ \\
  set all negative elements in $a$ to zero \\
  $stepNum = stepNum + 1$ \\
 }
 \Return{$a$}
}
\caption{SearchFromFree Algorithm}
\label{al:lisea}
\end{algorithm}

As shown in Algorithm 1, the \textbf{advGen} function has five inputs, including the local ML model $f'_{\theta'}$, and four positive constant parameters. The constant $step$ limits the maximum number of search iteration while $size$ defines the maximum modification of $a$ in each iteration. It is obvious that the  bill amount the energy thief from the utilities is approximately linear (considering the various electricity price) to $\left \| a \right \|_{1}$, the $L_1$-Norm of $a$. To maximize the attackers' profit, we add a regularization item $\lambda \cdot \left \| a \right \|_{1}$ to the loss $L_{total}$. We employ a constant $\lambda$ to represent the compromise between the attack success rate and the attacker's bill amount. Therefore, $\lambda$ should be carefully tuned according to specific attack scenarios. As shown by Line 5 of Algorithm 1, we empirically initialize $a$ according to a Gaussian distribution with the 
standard deviation value equals to $\sigma$ and mean value equals to zero, which indicates a zero amount (free) bill for the energy thief. The intuition behind this is that the search iteration process will gradually increase $\left \| a \right \|_{1}$ to a normal measurement vector that has a higher probability to bypass the detection. Therefore, a small initial $a$ will finally lead to a smaller $\left \| a \right \|_{1}$ so that the attacker can make more profit. The perturbation $r$ generated from the loss gradient may cause negative measurements in $a$. Therefore, as shown by Line 6 and 15, we set all negative values to zero to generate a feasible adversarial measurement vector $a$.

\section{Simulation Implementation} \label{sec:implementation}

\subsection{Dataset}

We utilize the smart meter data published by the Irish Social Science Data Archive (ISSDA) \cite{IrishData} as it is widely used as a benchmark for energy theft detection in related literature \cite{zanetti2017tunable}\cite{nabil2018deep, jokar2015electricity, korba2018energy}. The dataset contains the smart meter energy consumption measurement data of over 5000 customers in the Irish during 2009 and 2010. The customers agreed to install the smart meters and participated in the research project. Therefore, we assume all the measurement data in the dataset is normal and there is no energy theft. There are missing and illegal measurements in the original dataset. We pre-process the raw dataset by removing the incomplete measurements. We then regulate the time-series measurement data into daily reading vectors and obtain the dataset $D_{daily}$. Since the smart meters recorded every 30 minutes, each daily reading measurement vector will contain 48 measurements.

\begin{table}[htbp]
\caption{Energy Theft Attack Scenarios \cite{zanetti2017tunable}}
\begin{center}
\begin{tabular}{c}
\toprule[2pt]
\textbf{Attack Scenario}  \\
\midrule[1pt]
$h_{1}(m_{t}) = \alpha m_{t}$, $\alpha \sim Uniform(0.1, 0.8 )$ \\
\hline
$h_{2}(m_{t}) = \beta_{t} m_{t}$, $\beta_{t} \sim Uniform(0.1, 0.8 )$  \\
\hline
$h_{3}(m_{t}) = \left\{\begin{matrix}
0     & \forall t \in \left [ t_{i}, t_{f} \right ]\\ 
m_{t} & \forall t \notin \left [ t_{i}, t_{f} \right ]
\end{matrix}\right.$  \\
\hline
$h_{4}(m_{t}) = E(m)$  \\
\hline
$h_{5}(m_{t}) = \beta_{t} E(m)$ \\
\hline
$h_{6}(m_{t}) = m_{24-t}$\\
\bottomrule[2pt]
\end{tabular}
\end{center}
\label{table:scenario}
\end{table}

\begin{table*}[htbp]
\caption{Model Structures}
\begin{center}
\begin{tabular}{c|c|c|c|c|c|c}
% \midrule[1pt]
\toprule[2pt]
\textbf{Networks} & \multicolumn{2}{|c|}{\textbf{FNN}} & \multicolumn{2}{|c|}{\textbf{RNN}} & \multicolumn{2}{|c}{\textbf{CNN}} \\
\midrule[1pt]
\textbf{Models} & $f_{FNN}$ & $f'_{FNN}$ & $f_{RNN}$ & $f'_{RNN}$ & $f_{CNN}$ & $f'_{CNN}$  \\
\hline
\textbf{Layer 0} & input $48$ & input $48$ & input $48 \times 1$ & input $48 \times 1$ & input reshaped $6 \times 8$ & input  reshaped $6 \times 8$  \\
\hline
\textbf{Layer 1} & 128 Dense & 168 Dense & 256 LSTM & 246 LSTM & 128 Conv2D & 156 Conv2D \\
\hline
\textbf{Layer 2} & 256 Dense & 328 Dense & Dropout 0.25 & Dropout 0.25 & 128 Conv2D & 214 Conv2D \\
\hline
\textbf{Layer 3} & 128 Dense & 168 Dense & 168 LSTM & 148 LSTM & MaxPooling2D (2,2) & MaxPooling2D (2,2) \\
\hline
\textbf{Layer 4} & Dropout 0.25 & 128 Dense & Dropout 0.25 & Dropout 0.25 & Dropput 0.25 & Dropput 0.25\\
\hline
\textbf{Layer 5} & 32 Dense & Dropout 0.25 & 128 LSTM & 108 LSTM & flatten & flatten\\
\hline
\textbf{Layer 6} & Dropout 0.25 & 64 Dense & 2 Dense Softmax & 2 Dense Softmax & 32 Dense & 48 Dense \\
\hline
\textbf{Layer 7} & 2 Dense Softmax & Dropout 0.25 & - & - & Dense 2 Softmax & Dense 2 Softmax\\
\hline
\textbf{Layer 8} & - & 2 Dense Softmax & - & - & - & - \\
\bottomrule[2pt]
\multicolumn{7}{l}{The models $f_{\ast}$ act as the defenders while $f'_{\ast}$ as attackers. The activation function of each layer is $ReLu$ unless specifically noted.} \\
\multicolumn{7}{l}{The kernel size is $3 \times 3$ for all the CNN models.}
\end{tabular}
\end{center}
\label{table:struct}
\end{table*}

However, there is a shortage of real-world energy theft dataset. To solve this, we employ the false measurement data generation approach proposed in \cite{zanetti2017tunable} to simulate the energy theft measurements. As shown in Table \ref{table:scenario}, \cite{zanetti2017tunable} presents six energy theft scenarios. The first attack $h_{1}$ multiples the real meter reading with a constant while $h_{2}$ with a random constant generated from a uniform distribution. The $h_{3}$ assumes that the energy thief reports zero consumption during a period. The fourth scenario happens when an attacker constantly reports the mean consumption. $h_{5}$ is similar to $h_{2}$ but multiplying the random constant with the mean value instead of the real measurements. At last, $h_{6}$ reverses the records of a day so that the low measurements will be reported during the periods in which the electricity price is lower.

We generate a synthetic dataset based on the regulated daily smart meter dataset. We firstly randomly sample 180,000 daily records from $D_{daily}$ and pollute half records in the sampled dataset according to the attack scenarios described in Table \ref{table:scenario}. We label all normal records as 0 and polluted records as 1. We finally obtain a defender's dataset $D_{defender}:\left \{ M_{180,000 \times 48}, Y_{180,000 \times 1} \right \}$. We generate the dataset $D_{attacker}$ for the attacker in the same way. We note that the normal records in both $D_{attacker}$ and $D_{defender}$ are sampled from $D_{daily}$ and there are same records shared in the two datasets. However, since there are over 2.5 million records in $D_{daily}$, they can be regarded as different datasets.

\subsection{Evaluation}

We evaluate our algorithm with three kinds of neural networks, feed-forward Neural networks (FNN), RNN, and CNN. We train three neural networks for the utilities as the defender models and three separate neural networks with different structures for the attacker. The utilities' models and attacker's models are trained with $D_{defender}$ and $D_{attacker}$ respectively. The structures of the corresponding neural networks are shown in Table \ref{table:struct}.

\begin{table}[htbp]
\caption{Model Performance}
\begin{center}
\begin{tabular}{c|c|c}
\toprule[2pt]
\textbf{Model} & \textbf{Accuracy} & \textbf{False Positive Rate} \\
\midrule[1pt]
$f_{FNN}$ & 86.9\% & 10.01\% \\
\hline
$f'_{FNN}$ & 86.87\% & 14.01\% \\
\hline
$f_{RNN}$ & 97.5\% & 2.58\% \\
\hline
$f'_{RNN}$ & 97.48\% & 2.62\% \\
\hline
$f_{CNN}$ & 93.49\% & 7.79\% \\
\hline
$f'_{CNN}$ & 93.28\% & 6.41\% \\
\bottomrule[2pt]
\end{tabular}
\end{center}
\label{table:NNperf}
\end{table}

All our ML models are implemented with the TensorFlow and Keras library. We conduct our experiments on a Windows 10 PC with an Intel Core i7 CPU and 16 GB memory. An NVIDIA GeForce GTX 1070 graphic card is employed to accelerate the training process. The related source code of this paper is available on Github \cite{SearchFromFree}.

We manually tuned the parameters of the model training and the performances of different models are shown in Table \ref{table:NNperf}. From the table, we can learn that the RNN (LSTM) achieves the best performance in both metrics, followed by the CNN and FNN. This is because RNNs have an inherent advantage in learning the pattern of time-series data. Meanwhile, we can also learn that the performances of attackers' models are very similar to the defenders' models, which indicates that our sampled datasets can represent the overall manifold of the raw dataset.

\noindent \textbf{Evaluation Metrics: } We set two metrics to evaluate the performance of our attacks. The first metric will be the detection accuracy of the defender's ML models under the adversarial attacks. Meanwhile, it is straightforward that a fake measurement vector with a smaller profit to the energy thief will have a higher probability to bypass the energy theft detection. For example, if the attacker just sets the parameter $\alpha$ of $h_{1}$ in Table \ref{table:scenario} to $1-10^{-5}$, the adversarial measurement vector will hold a very high probability to be classified as normal by the utilities' ML model. However, this will lead to a smaller profit of the attacker. Therefore, we will select the average $L_{1}$-Norm of the fake measurement vectors as the second evaluation metric. In our experiment, the average $L_{1}$-Norm of all normal measurement records in $D_{defender}$ is 32.05 kWh.

\begin{figure}[htbp]
\centerline{\includegraphics[width=1\linewidth]{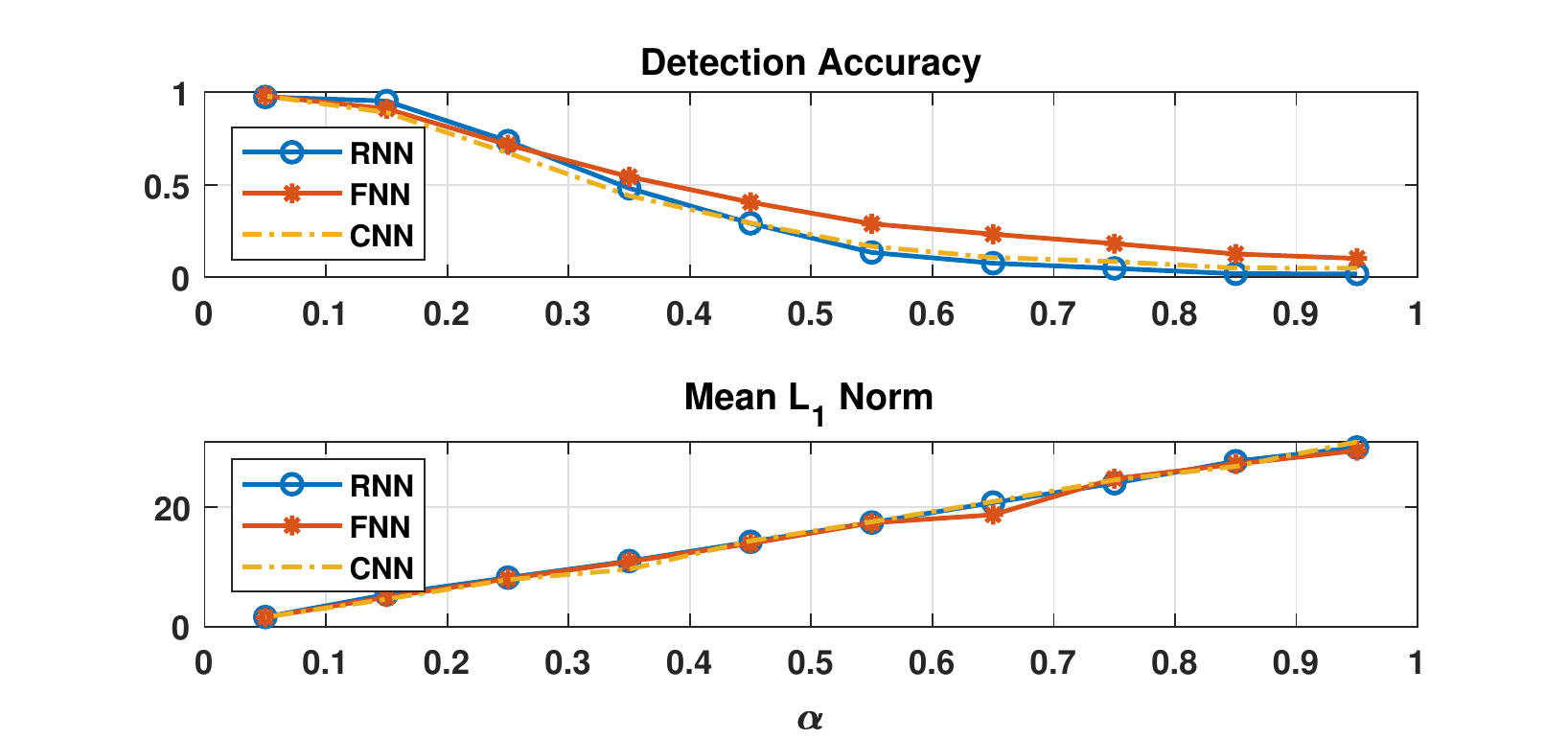}}
\caption{Vanilla Attacker 1}
\label{fig:vanilla1}
\end{figure}

\begin{figure}[htbp]
\centerline{\includegraphics[width=1\linewidth]{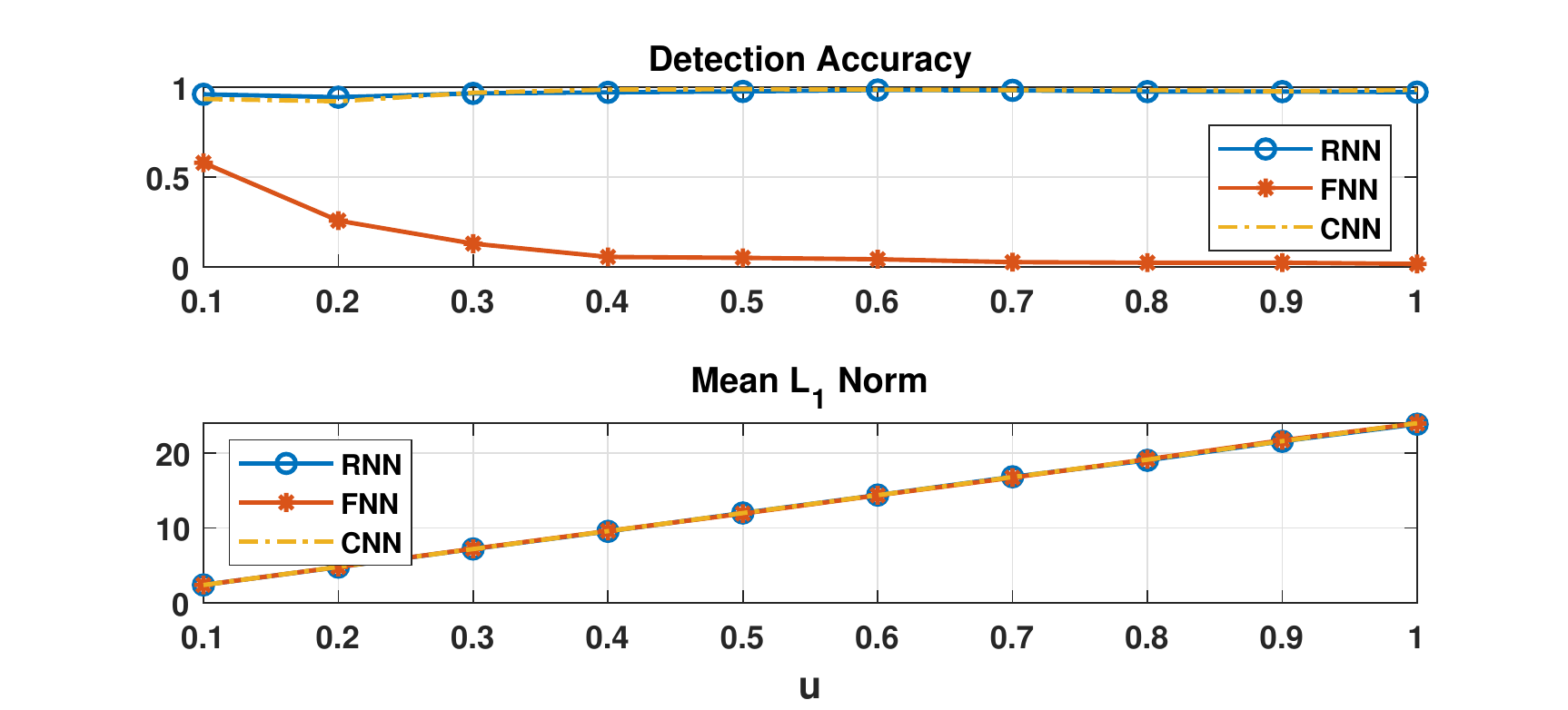}}
\caption{Vanilla Attacker 2}
\label{fig:vanilla2}
\end{figure}

\noindent \textbf{Baselines: } We set up two \textbf{vanilla black-box attackers} as baselines. The first vanilla attacker $\textbf{VA1}$ will simply gradually try different $\alpha$ of $h_{1}$ as defined in Table \ref{table:scenario} while the second vanilla attacker $\textbf{VA2}$ generates uniformly distributed measurement vector between 0 and a variable $u$. The performances of the vanilla attackers with 1,000 fake measurement vectors are demonstrated in Fig. \ref{fig:vanilla1} and Fig. \ref{fig:vanilla2} respectively. 

From Fig. \ref{fig:vanilla1}, we can learn that the detection accuracy of the defenders' models decreases with the parameter $\alpha$ increases under $\textbf{VA1}$ attack. As analyzed above, this indicates that the energy thief has a higher success probability if the he/she was willing to decrease his/her stolen profit. Overall, CNN and RNN are more vulnerable to $\textbf{VA1}$ attack compared with FNN. This is because the RNN and CNN can learn the time-series pattern more accurately than FNN. However, by comparing the two figures in Fig. \ref{fig:vanilla1}, if the attacker wants to have a relatively high success probability for energy theft, such as over 65\%, she/he will need to pay an over 20 kWh power consumption bill ($\alpha > 0.65$).

The performance of the $\textbf{VA2}$ attack is demonstrated in Fig. \ref{fig:vanilla2}. We can learn that the detection accuracy of RNN and CNN remains high (over 95\%) with the parameter $u$ increases. Again, this verifies our analysis that the RNN and CNN can learn the daily electricity consumption patterns, and a uniformly distributed consumption measurement vector is obviously abnormal. However, the performance of the FNN decreases dramatically if the attacker generates larger fake measurements. Overall, the $\textbf{VA2}$ attack is not effective for energy theft for the attacker.

\begin{figure}[htbp]
\centerline{\includegraphics[width=1\linewidth]{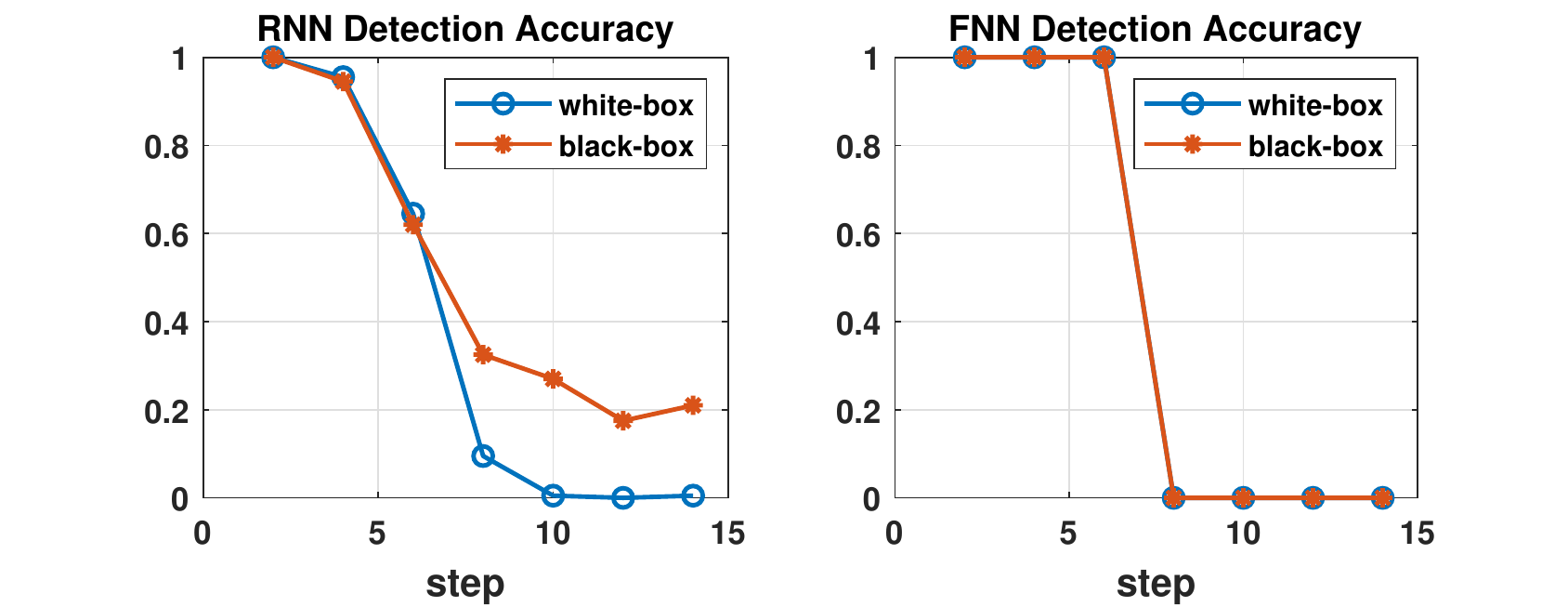}}
\caption{The detection accuracy of the RNNs and the FNNs. The parameter $size=0.01$ while $\lambda = 10$. }
\label{fig:RNN-FNN-Accu}
\end{figure}

\begin{figure}[htbp]
\centerline{\includegraphics[width=1\linewidth]{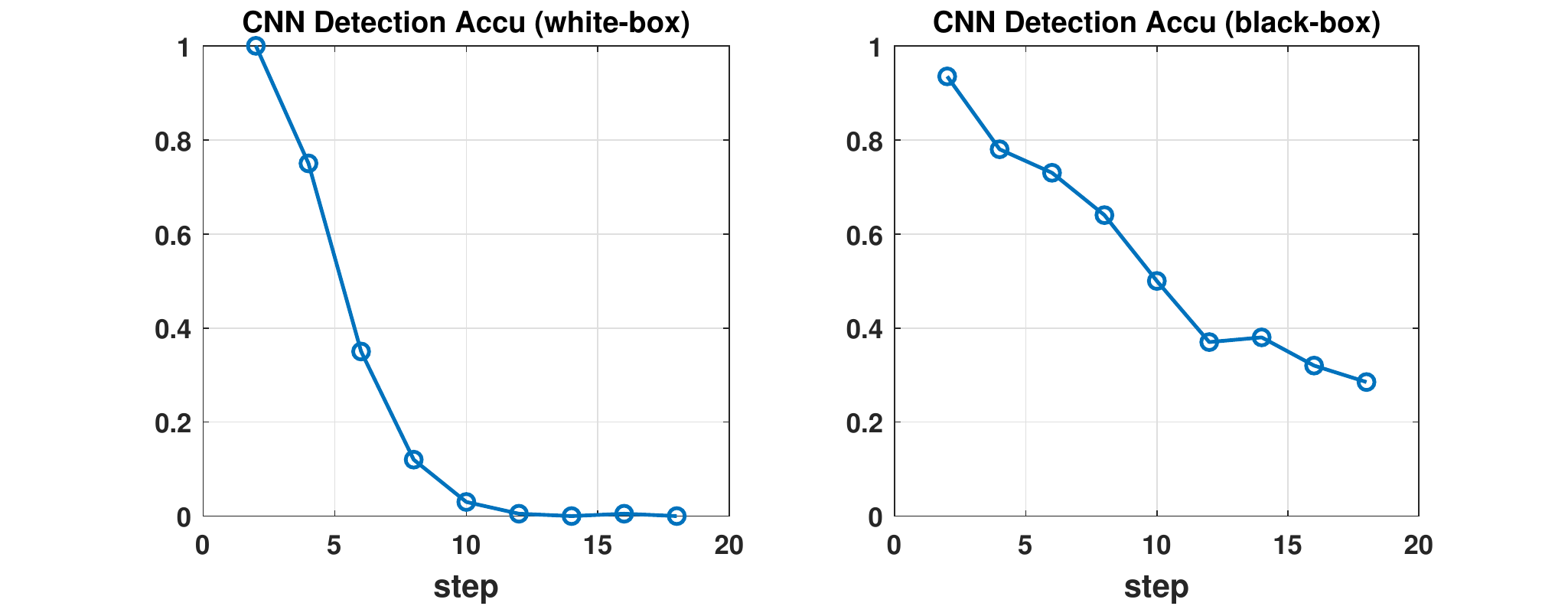}}
\caption{The detection accuracy of the CNNs. The parameter $\lambda = 10$ in white-box attacks and $\lambda = 0$ in black-box attacks. The white-box attack has a $size=0.01$ while the black-box attack sets $size=0.1$.}
\label{fig:CNN-Accu}
\end{figure}

We then evaluate the effectiveness of the \textit{SearchFromFree} algorithm. We consider two types of attack, the \textbf{white-box} attack and \textbf{black-box} attack. The \textbf{white-box} attacker is assumed to have full access to the utilities' models $f_{\ast}$ while the \textbf{black-box} attacker can only access to his/her alternative models $f'_{\ast}$. Fig. \ref{fig:RNN-FNN-Accu} and Fig. \ref{fig:CNN-Accu} demonstrate the energy theft detection accuracy of different ML models under the white-box attacks and black-box attacks. The experiments were also conducted with 1,000 generated adversarial measurement vectors per case. From the left figure of Fig. \ref{fig:RNN-FNN-Accu}, we can learn that the detection accuracy drops dramatically with the $step$ increases. When $step$ becomes larger, the RNN models' detection accuracy is close to zero under the white-box attacks and is around 20\% under the black-box attacks. The FNN's detection accuracy is even worse under our adversarial attacks. The right figure in Fig. \ref{fig:RNN-FNN-Accu} shows that the FNN models have no energy theft detection ability under both white-box and black-box attacks.

Figure \ref{fig:CNN-Accu} presents the detection accuracy of the CNN models under the \textit{SearchFromFree} attacks. Similar to RNN and FNN, the detection accuracy of the CNN model drops to zero under white-box attacks. The accuracy is slightly higher in the black-box scenario (around 30\%) but is still far from reliability. Meanwhile, a larger search $size$ is needed to achieve a good attack performance for the black-box CNN attack in our experiments.

\begin{figure}[htbp]
\centerline{\includegraphics[width=1\linewidth]{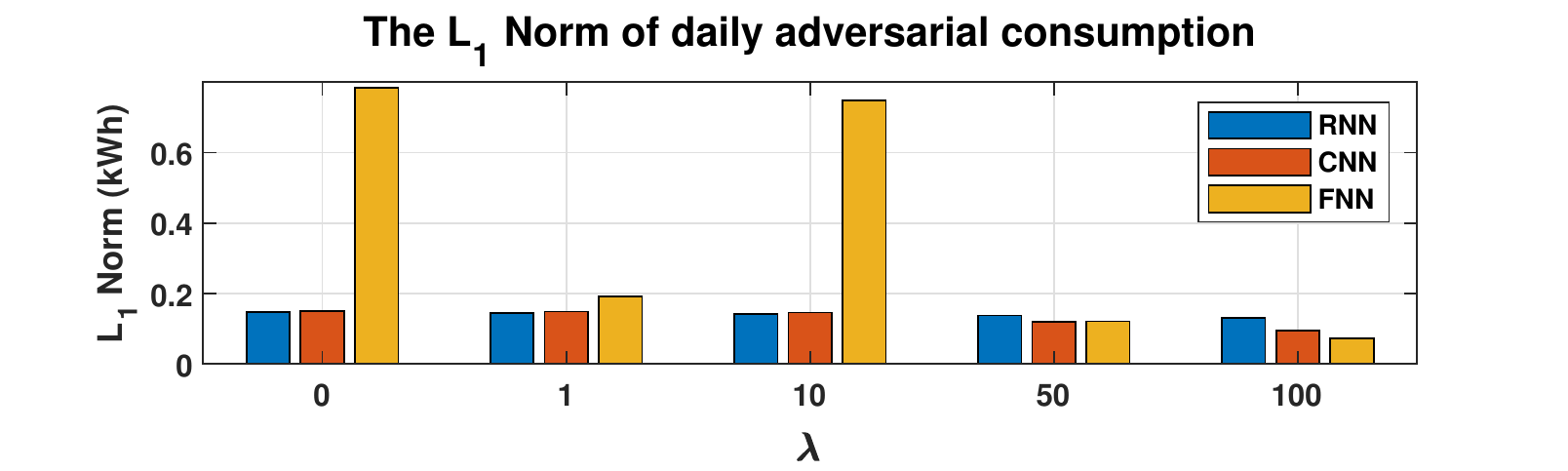}}
\caption{The average $L_{1}$-Norm of adversarial measurement vectors (white-box attacks). The parameters $size = 0.01$ and $step = 14$ for all three models.}
\label{fig:NormWhite}
\end{figure}

\begin{figure}[htbp]
\centerline{\includegraphics[width=1\linewidth]{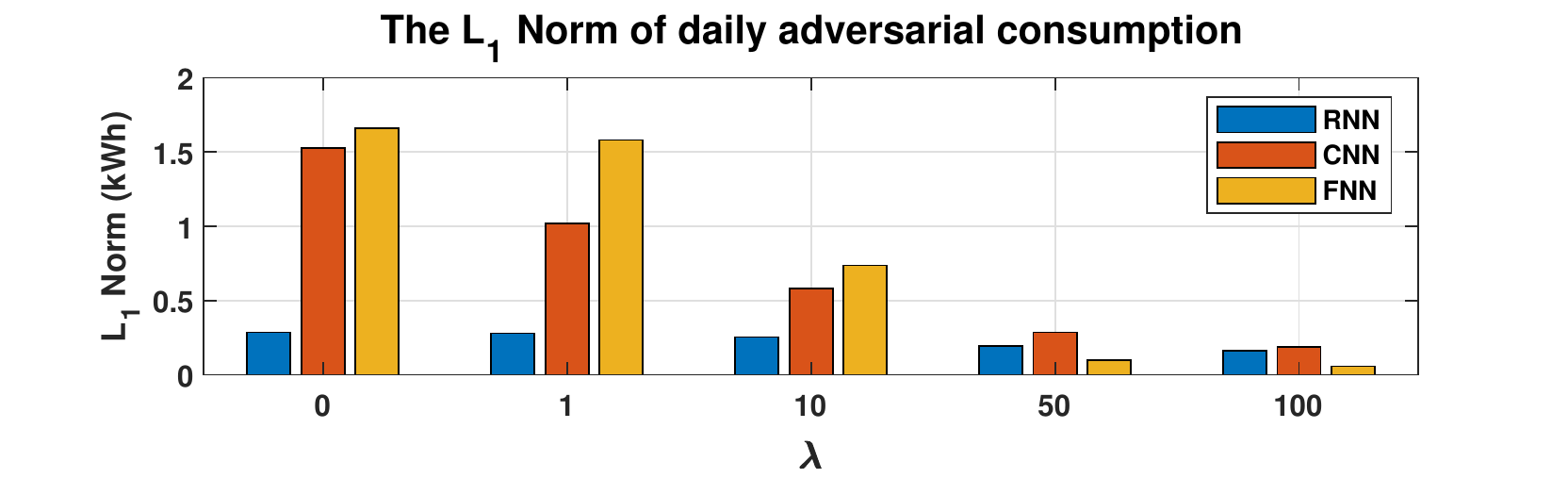}}
\caption{The average $L_{1}$-Norm of adversarial measurement vectors (black-box attacks). The parameter $step = 14$ for all RNN and FNN and $step = 18$ for CNN. For RNN and FNN, $size$ is set to 0.01 while for CNN is 0.1.}
\label{fig:NormBlack}
\end{figure}

The values of the average $L_{1}$-Norm of the corresponding adversarial measurement vectors under the white-box attacks and the black-box attacks are illustrated in Fig. \ref{fig:NormWhite} and Fig. \ref{fig:NormBlack} respectively. From Fig. \ref{fig:NormWhite}, we can learn that the compromise parameter $\lambda$ has a larger effect on the FNN than the RNN and the CNN. On the other hand, as demonstrated in Fig. \ref{fig:NormBlack}, the average $L_{1}$-Norm decreases for all models while $\lambda$ increases. In our experiments, we find that a large $\lambda$ (larger than 10) will significantly decrease the attacks' performance on bypassing the detection. Therefore, the attacker may prefer a relatively small $\lambda$ to increase his/her overall stolen profit in practical scenarios. 

By comparing Fig. \ref{fig:NormWhite} and Fig. \ref{fig:NormBlack} with Fig. \ref{fig:RNN-FNN-Accu} and Fig. \ref{fig:CNN-Accu}, we can summarize the performances of our attacks in Table \ref{table:perSum}. The worst attack performance in our experiments is from the black-box attack of $f'_{CNN}$ (31\% Accu, 1.59 kWh), which is still effective enough for energy theft, compared with the vanilla attacks and the 32.05 kWh normal daily consumption.

\begin{table}[htbp]
\caption{Attack Performance Summary}
\begin{center}
\begin{tabular}{c|c|c|c|c}
\toprule[2pt]
\textbf{Attack} & \textbf{Model} & \textbf{Accu} & \textbf{$L_1$-Norm} & \textbf{Parameters} \\
\midrule[1pt]
\multirow{3}*{White-box} & $f_{RNN}$ & 0.5\% & 0.17 & $Size = 0.01$, $\lambda = 10$\\
 \cline{2-5}
  ~  & $f_{CNN}$ & 0.2\% & 0.18 & $Size = 0.01$, $\lambda = 10$ \\
 \cline{2-5}
 ~ & $f_{FNN}$ & 0\% & 0.76 & $Size = 0.01$, $\lambda = 10$ \\
 \hline
 \multirow{3}*{Black-box} & $f'_{RNN}$ & 21\% & 0.28 & $Size = 0.01$, $\lambda = 10$\\
 \cline{2-5}
  & $f'_{CNN}$ & 31\% &  1.59 &  $Size = 0.1$, $\lambda = 0$\\
 \cline{2-5}
  & $f'_{FNN}$ & 0\% & 0.75 & $Size = 0.1$, $\lambda = 10$\\
\bottomrule[2pt]
 \multicolumn{5}{l}{$^{\ast}$ The $step = 18$ for $f'_{CNN}$ and $step = 14$ for the rest.}
\end{tabular}
\end{center}
\label{table:perSum}
\end{table}

\section{Discussion and Future Work} \label{sec:discussion}

We investigate the vulnerability of the ML-based energy theft detection from the attacker's perspective. A more reliable ML solution that is resistant to the adversarial attacks needs to be studied in the future. The state-of-the-art defense approaches, such as adversarial training, will be evaluated in our future work. On the other hand, energy theft is inevitable even if ML-based detection is employed since the energy thieves can just modify the $h_{1}$ attack defined in Table \ref{table:scenario} with a relatively larger $\alpha$. Therefore, defense mechanisms that can significantly decrease the energy theft profit of the attackers should also be investigated.

\section{Conclusion} \label{sec:conclusion}

The ML-based energy theft detection mechanisms are highly vulnerable to well-designed adversarial attacks. In this paper, we study the adversarial attacks in energy theft detection and propose a general threat model. We then design the \textit{SearchFromFree} algorithm to maximize the attacker's profit. We evaluate our attacks with three kinds of neural networks trained on a real-world smart meter dataset. The result demonstrates that the energy thieves can effectively bypass the detection of the well-trained ML models with extremely low reported energy consumption measurements, even in black-box attack scenarios.

\bibliography{reference} 
\bibliographystyle{ieeetr}

\end{document}